\documentclass{article}
\usepackage{spconf,amsmath,graphicx}
\usepackage{multirow}
\usepackage[psamsfonts]{amssymb}
\usepackage{amsxtra}
\usepackage{threeparttable}
\usepackage{cite}

\def\x{{\mathbf x}}

\usepackage{color}
\title{Multi-channel Speech Separation Using Deep Embedding Model with Multilayer Bootstrap Networks
}
\name{Ziye Yang$^1$ and Xiao-Lei Zhang$^{1,2}$}
\address{$^1$Center for Intelligent Acoustics and Immersive Communications and\\
  School of Marine Science and Technology, Northwestern Polytechnical University, China\\
  $^2$Research \& Development Institute of Northwestern Polytechnical University in Shenzhen, China \\
 2015300797@mail.nwpu.edu.cn, xiaolei.zhang@nwpu.edu.cn\\
\thanks{
This work was supported in part by the National Natural Science Foundation of China (NSFC) funding scheme under Project No. 61671381, in part by the Project of the Science, Technology, and Innovation Commission of Shenzhen Municipality under grant No. JCYJ20170815161820095, and in part by the Shaanxi Natural Science Basic Research Program under grant No. 2018JM6035.
}}
\begin{document}
\ninept
\maketitle
\begin{abstract}
Recently, deep clustering (DPCL) based speaker-independent speech separation has drawn much attention, since it needs little speaker prior information. However, it still has much room of improvement, particularly in reverberant environments. If the training and test environments mismatch which is a common case, the embedding vectors produced by DPCL may contain much noise and many small variations. To deal with the problem, we propose a variant of DPCL, named DPCL++, by applying a recent unsupervised deep learning method---\textit{multilayer bootstrap networks} (MBN)---to further reduce the noise and small variations of the embedding vectors in an unsupervised way in the test stage, which fascinates \textit{k}-means to produce a good result. MBN builds a gradually narrowed network from bottom-up via a stack of \textit{k}-centroids clustering ensembles, where the \textit{k}-centroids clusterings are trained independently by random sampling and one-nearest-neighbor optimization. To further improve the robustness of DPCL++ in reverberant environments, we take spatial features as part of its input.
Experimental results demonstrate the effectiveness of the proposed method.
\end{abstract}
\begin{keywords}
cocktail party problem, speaker-independent speech separation, deep clustering, multilayer bootstrap networks
\end{keywords}
\section{INTRODUCTION}
\label{sec:intro}
Speech separation is the task of separating target speech from interference background \cite{Wang2017Supervised}. According to the number of microphones, speech separation can be divided into single channel speech separation and multiple channel speech separation \cite{bregman1994auditory}. According to whether speakers' information is predefined or known as a prior, speech separation can be divided into three categories---speaker-dependent\cite{bregman1994auditory}, target-dependent\cite{zhang2016deep}, and speaker-independent \cite{yu2017permutation,isik2016single,hershey2016deep} speech separation.
Traditional speech separation methods include computational auditory scene analysis \cite{Wang2008Computational,Shao2005Model}, non-negative matrix factorization \cite{Lyubimov2013Non}, and minimum mean square error \cite{Ephraim2003Speech}. Recently, deep-learning-based supervised speech separation has attracted much attention as a new research trend \cite{Wang2017Supervised}. This paper focuses on deep-learning-based speaker-independent speech separation \cite{yu2017permutation,isik2016single,hershey2016deep}, since it does not require speaker identities in the test stage.

Deep-learning-based speaker-independent speech separation can be roughly categorized into three classes. The first class is deep clustering (DPCL)\cite{hershey2016deep,isik2016single,chen2017deep}. It generates an embedding vector for each time-frequency unit of a mixed magnitude spectrum by minimizing the Frobenius norm between the affinity matrix of the embedding vectors and the affinity matrix assigned by the ideal speakers. Bi-directional long short-term memory networks (BLSTM) is usually adopted as the deep learning toolbox for producing the embedding vectors. The second class is permutation invariant training (PIT)\cite{yu2017permutation,Kolb2017Multi}. It calculates the local mean
squared errors of all permutations of training speakers at either the frame-level or the utterance-level, and pick the locally optimal permutation corresponding to the minimum mean squared error to train the separation network. The third type is end-to-end speech separation \cite{Yi2017TasNet,Luo2018TasNet,Venkataramani2017End,Shi2019FurcaNet,Shi2019FurcaNeXt}. It builds models on time domain speech directly using an encoder-decoder framework and performs the source separation on nonnegative encoder outputs. Although these methods work well in clean environments, their performance degrades significantly in reverberant environments.

 To improve the performance of speech separation in reverberant environments, many multichannel methods based on DPCL were proposed. They can be mainly categorized into two classes---beamforming \cite{higuchi2017deep} and spatial feature extraction \cite{wang2018multi,jiang2014binaural,araki2015exploring,pertila2015distant}. The first class predicts a mask for each speaker at each channel by DPCL, and then conducts beamforming for each speaker by applying the masks of the speaker to estimate the beamforming coefficients, where the beamformers include the maximum signal-to-noise ratio beamformer \cite{higuchi2017deep} and minimum variance-distortion-free response beamformer \cite{griffiths1982alternative,warsitz2007blind}. The second class combines spatial features and spectral features together for the DPCL training.
 
This paper pursues DPCL, since it demonstrates good performance in many challenging scenarios. One weakness of DPCL is that it uses a clustering algorithm to partition the embedding vectors into different speakers. Because the BLSTM model of DPCL is trained in a supervised way, the embedding vectors contain the mismatching information between the training and test, such as random noise and small variations. It is known that clustering methods are sensitive to random noises, particularly in the cases where the clustering methods themselves suffer from some weaknesses, such as bad local minima and prior assumptions. The weaknesses of the \textit{k}-means clustering of DPCL may significantly degrade the performance in reverberant scenarios. Although some work replaced the \textit{k}-means clustering by a PIT-based clustering \cite{Fan2019Discriminative}, the PIT network still needs supervised training.

 In this paper, we propose to reduce the random noise and small variations of the embedding vectors by a recently proposed unsupervised deep model, named \textit{multilayer bootstrap networks} (MBN).
MBN is a simple nonlinear dimensionality reduction method \cite{Zhang2018Multilayer}. It builds a gradually narrowed multilayer network by a stack of $k$-centroids clustering ensemble without resorting to neural network architectures. Each $k$-centroids clustering is trained by random sampling of data and one-nearest-neighbor optimization. MBN does not make data and model assumptions, and does not suffer the weaknesses of neural networks. MBN provides clean data representations with little random noise and small variations, which helps the \textit{k}-means clustering of DPCL suffer less from its weaknesses. To further deal with reverberant environments, we extract a spatial feature, named cosine interchannel
phase difference (cosIPD) as part of the input of DPCL. We name the overall system as DPCL++. Experimental results demonstrate the effectiveness of the proposed method.


\section{SYSTEM DESCRIPTION}

Figure 1 shows an overview of the proposed DPCL++ system. It contains three components---feature extractor, deep clustering, and MBN, which will be presented in Sections \ref{subsec:spatial} to \ref{subsec:mbn} respectively.

\begin{figure}[t]
 \centering
  \includegraphics[width=0.32\textwidth]{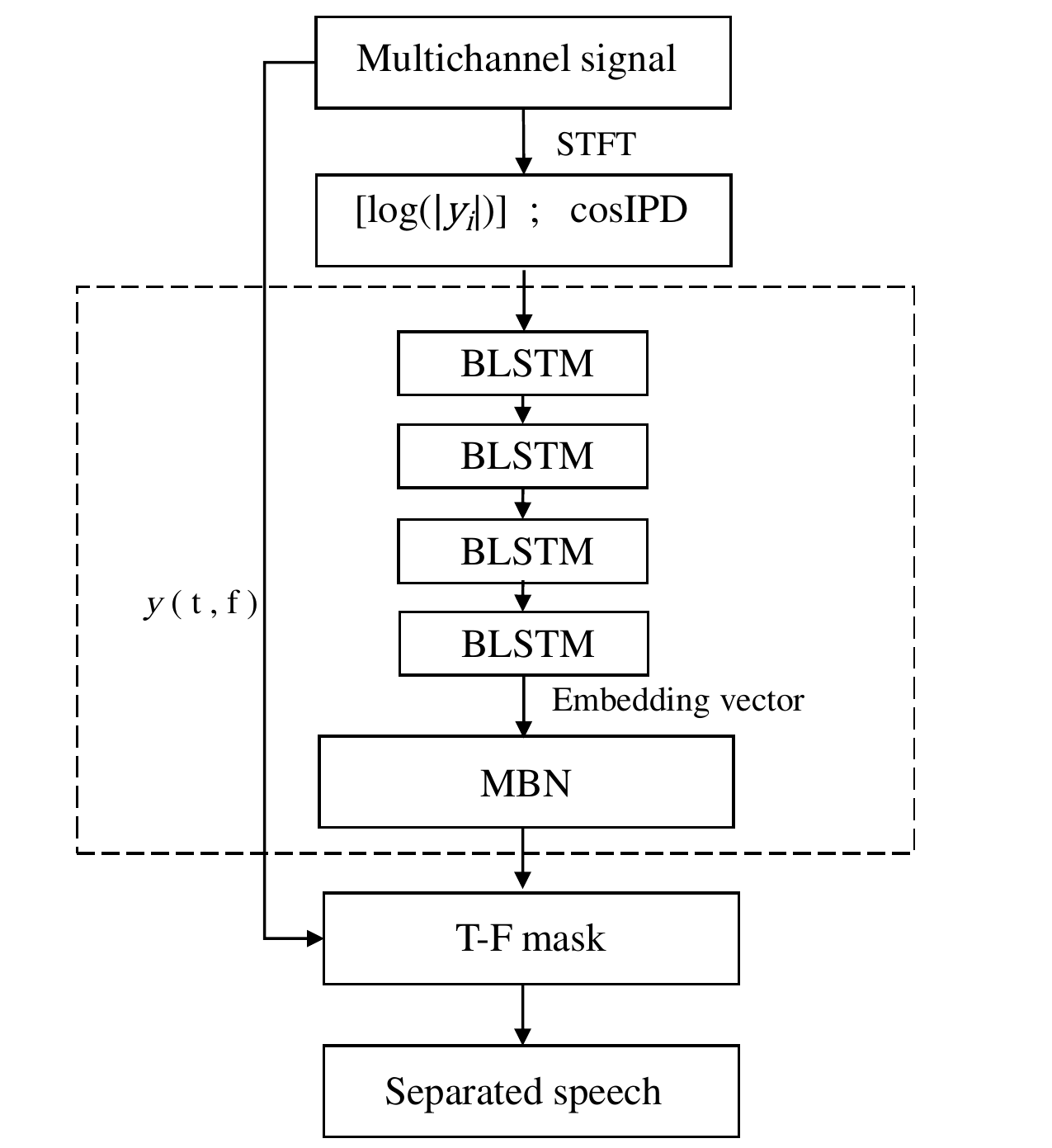}
  \caption{Diagram of the proposed DPCL++ system.}
  \end{figure}

%

\subsection{Feature extraction}\label{subsec:spatial}

In the training stage, we first extract $2$ short time Fourier transform STFT spectrograms from each of the two audio recordings, denoted as $\{y_{i,1},y_{i,2}\}_{i=1}^n$, where $i$ is a time-frequency (T-F) index $(t,f)$ at time $t$ and frequency $f$, $n$ is the total number of the T-F units of a STFT spectrogram, and $y_{i,p}$ denotes the $i$-th T-F unit of the $p$-th spectrogram with $p \in\{ 1,2\}$. Then, we extract a log-magnitude spectrum $ \log \left|y_{i,p}\right|$ and a spatial feature interchannel phase difference $\angle y_{i,1}-\angle y_{i,2}$.
To handle the $2\pi$ ambiguity, we further transform IPD by a cosine function, i.e. $\cos(\angle y_{i,1}-\angle y_{i,2})$,  so as to unwrap
the phase values into a range $[-1,1]$ \cite{wang2018multi}.
Finally, the input acoustic feature of the $i$-th T-F unit is:
\begin{equation}\label{eq:x}
\mathbf{z}_{i} = \left[ \log \left|y_{i,1}\right|, \log \left|y_{i,2}\right|,\cos(\angle y_{i,1}-\angle y_{i,2}) \right]^T
 \end{equation}

\subsection{Deep clustering}\label{subsec:deep}

DPCL++ learns a $k$-dimensional embedding vector $\mathbf{x}_{i}$ for $\mathbf{z}_{i}$ by a BLSTM network $g(\cdot)$:
$\mathbf{x}_{i} = g(\mathbf{z}_{i})$.
The BLSTM network minimizes the following cost function:
  \begin{equation}\label{eq:blstm}
\mathcal{J}=||\mathbf{X}^T\mathbf{X}-\mathbf{B}^T\mathbf{B}||_F^2
 \end{equation}
where $\|\cdot\|_F$ denotes the Frobenius norm operator, $\mathbf{X} = [\mathbf{x}_{1},\ldots, \mathbf{x}_{n}]$ is an $n\times k$ embedding matrix, and $\mathbf{B} = [\mathbf{b}_{1},\ldots, \mathbf{b}_{n}]$ is an $n \times U$ ground-truth indicator matrix with $\mathbf{b}_{i} = [b_{i,1},\ldots,b_{i,u},\ldots,b_{i,U}]^T$ defined as:
  \begin{equation}\label{eq:blstm}
b_{i,u} = \left\{\begin{array}{ll}
1,& \mbox{ } \mbox{if the T-F unit is dominated by speaker $u$.}\\
0,& \mbox{ } \mbox{otherwise.}
\end{array}.\right.
 \end{equation}

In the test stage, suppose $O$ speakers talk simultaneously. We first use MBN to transform the embedding vectors $\mathbf{x}$ to a new feature representation, named m-vectors $\mathbf{m}$, and then use the \textit{k}-means clustering to partition $\mathbf{m}$ into $O$ clusters, which generates $O$ estimated binary masks, each of which for a speaker:
 \setlength{\arraycolsep}{0.0em}
  \begin{eqnarray}\label{eq:ebm}
&\hat{M}_{o}(t,f) = \left\{\begin{array}{ll}
1,&  \mbox{if the $(t,f)$-unit is assigned to speaker $o$.}\\
0,& \mbox{otherwise.}
\end{array}\right.,\nonumber \\
&\forall o = 1,\ldots, O.
 \end{eqnarray}

\subsection{Multilayer bootstrap networks}\label{subsec:mbn}

\subsubsection{Method}

\begin{figure}[t]
 \centering
  \includegraphics[width=0.3\textwidth]{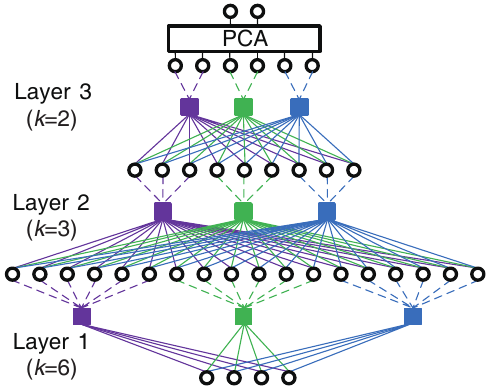}
  \caption{ Network structure of MBN \cite{Zhang2018Multilayer}. The dimension of the input data for this demo network is 4. Each colored square represents a k-centroids clustering. Each layer contains 3 clusterings. Parameters k at layers 1, 2, and 3 are set to 6, 3, and 2 respectively. The outputs of all clusterings in a layer are concatenated as the input of their upper layer.
 }
 \label{fig:mbn}
  \end{figure}
  
  MBN is a recently proposed nonlinear dimensionality reduction method. 
As illustrated in Fig. \ref{fig:mbn}, it has multiple hidden layers and an output layer. Each hidden layer consists of $V$ independent \textit{k}-centroids clusterings, where $V\gg 1$. Each \textit{k}-centroids clustering has \textit{k} output units, each of which indicates a cluster. The output units of all \textit{k}-centroids clusterings in the same layer are concatenated as the input of their upper layer. The output layer is principal component analysis (PCA).

MBN is built layer-by-layer from bottom-up as a gradually narrowed network. Suppose MBN contains $L$ layers, and the parameters $k$ from the bottom hidden layer to the top hidden layer are denoted as $k_1,\ldots, k_L$ respectively. The parameters $k_1,\ldots, k_L$ are determined by the following criteria:
\begin{eqnarray}
 &k_1 \gg O, \label{eq:k1}\\ 
 & k_{l+1}=\delta k_l,\label{eq:rel}\quad \forall l = 1,\ldots,L-1,\\ 
 & k_L \mbox{is set to eusure at least one data point per class in probability}\label{eq:kL}
\end{eqnarray}
where $k_1$ and $\delta\in [0,1)$ are user-defined hyperparameters. It can be seen that $L$ is determined automatically. Note that the criterion \eqref{eq:kL} is usually specified to $k_L \geq  \left\lceil 1.5 O\right\rceil$ for class-balanced problems.

 
 For training each layer given a $d$-dimensional input data set $\mathcal{X} = \left\{\mathbf{x}_1,\ldots,\mathbf{x}_n \right\}$ either from the lower layer or from the output of the BLSTM model, MBN trains each $k$-centroids clustering independently via the following steps \cite{Zhang2018Multilayer}:
 \begin{itemize}
 \itemsep=0.0pt
   \item \textbf{Random sampling of features.} The first step randomly selects $\hat{d}$ dimensions of $\mathcal{X}$ ($\hat{d}\le d$) to form a subset of $\mathcal{X}$, denoted as $\hat{\mathcal{X}} = \left\{\hat{\mathbf{x}}_1,\ldots,\hat{\mathbf{x}}_n \right\}$.
    \item \textbf{Random sampling of data.} The second step randomly selects $k$ data points from $\hat{{\mathcal{X}}}$ as the $k$ centroids of the clustering, denoted as $\{\mathbf{w}_{1},\ldots,\mathbf{w}_{k} \}$.
   \item \textbf{One-nearest-neighbor learning.} The new representation of an input $\hat{\x}$ produced by the current clustering is an indicator vector $\mathbf{h}$ which indicates the nearest centroid of $\hat{\x}$. For example, if the third centroid is the nearest one to $\hat{\mathbf{x}}$, then $\mathbf{h} = [0,0,1,0,\ldots,0]^T$. The similarity metric between the centroids and $\hat{\mathbf{x}}$ at the bottom layer is the squared Euclidean distance $ \arg\min_{i=1}^{k}\|\mathbf{w}_i-\hat{\mathbf{x}}\|^2$, and set to $\arg\max_{i=1}^{k}\mathbf{w}_i^T\hat{\mathbf{x}}$ at all other hidden layers.
 \end{itemize}

\subsubsection{Fundamentals}

\begin{figure}[t]
\begin{minipage}[t]{0.5\linewidth}
\centering

\includegraphics[width=40mm]{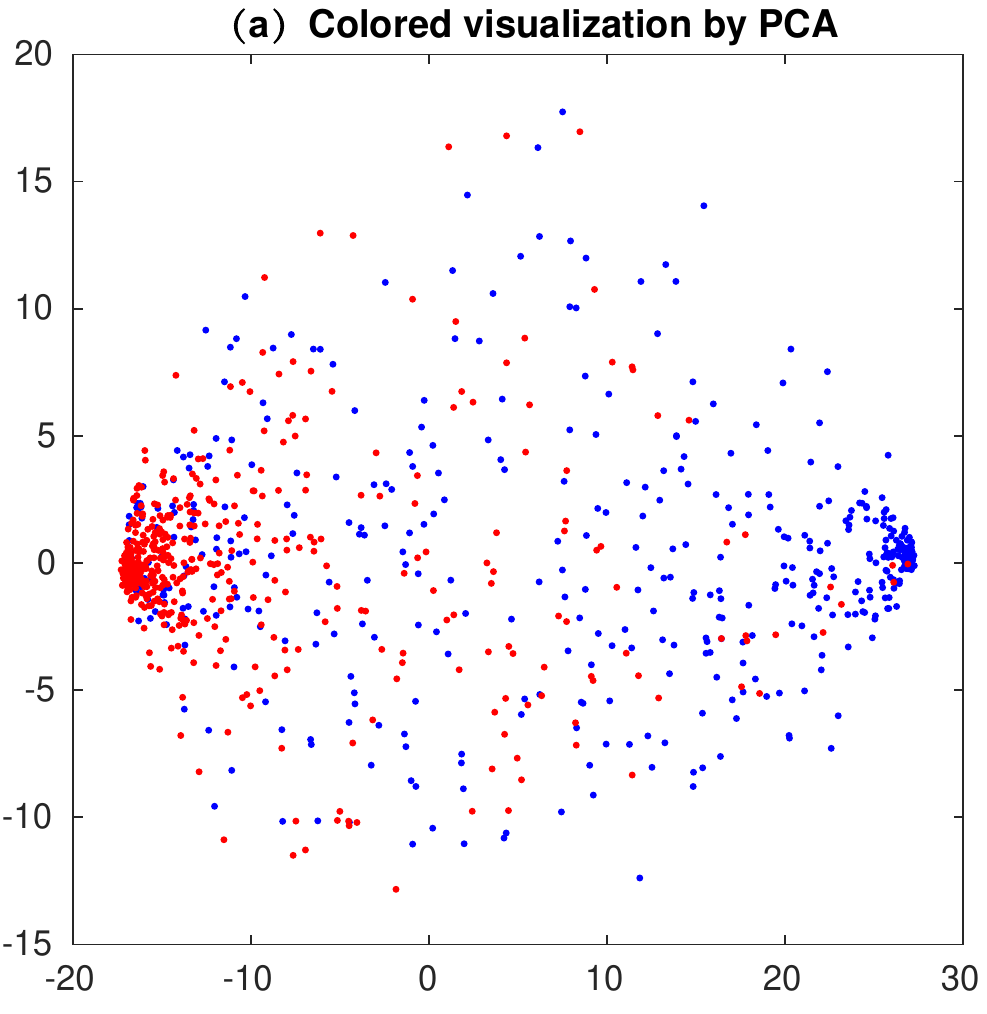}
  \label{fig:side:b}
\end{minipage}%
\begin{minipage}[t]{0.5\linewidth}
\centering
\includegraphics[width=42mm]{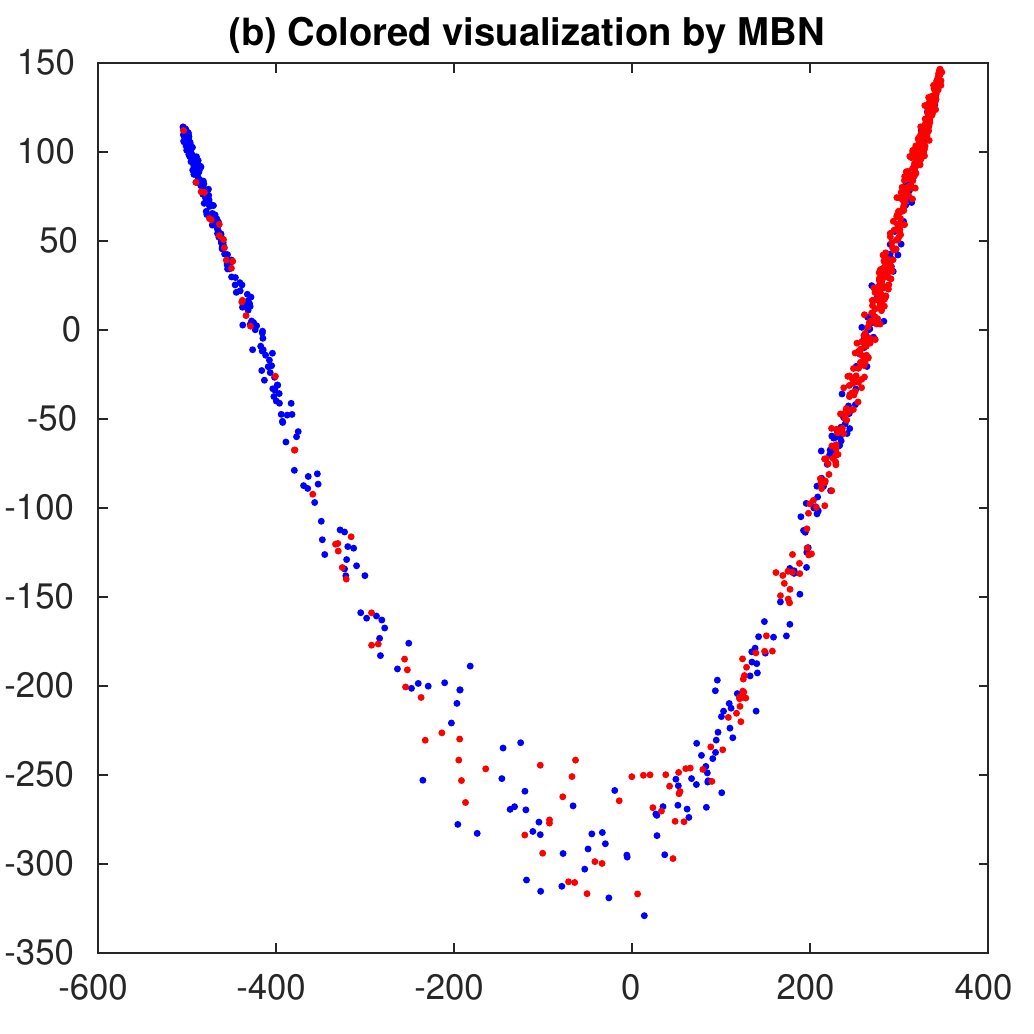}
  \label{fig:side:a}
 \end{minipage}

   \caption{Visualizations of (a) the embedding vectors produced by BLSTM and (b) the $m$-vectors produced by MBN. }
 \label{fig:PCAvsMBN}
\end{figure}

 MBN has four hyperparameters $\{V, a, k_1, \delta\}$. We present their fundamentals and selection criteria respectively as follows.
 
 First, MBN is an unsupervised deep ensemble learning method. Like many ensemble learning methods such as bootstrap aggregating, the estimation error of MBN is decreased linearly along with the increase of the number of the base clusterings, i.e. $V$, and $V$ is usually set to a number of a few hundreds. 
 
 Second, decorrelating the base clusterings by random resampling of features also enlarges the diversity between the base clusterings, as analyzed in random forests. However, because the embedding vectors are produced by BLSTM, their dimensions are not mutually independent. In other words, they may contain similar information. Hence, random resampling of the features may not yield large performance improvement. On the contrary, setting $a$ to a small value may cost an inaccurate estimate of the distribution of the input data at each base clustering. Therefore, we set $a$ to a number close to 1 for DPCL++.
 
 Third, $k_1$ is a very important hyperparameter that balances the estimation accuracy and computational complexity. Specifically, MBN was originally designed for the dimensionality reduction problem of highly nonlinear data. It has a fundamental assumption that, for any $k$-centroids clustering, the small area around a centroid of the clustering is locally linear.  The larger $k_1$ is set to, the more likely MBN captures the small nonlinear variations of data. However, the computational complexity is also increased linearly with respect to the increase of $k_1$. As shown in Fig. \ref{fig:PCAvsMBN}a, the discriminability of the embedding vectors is affected mainly by random noise; the nonlinearity has been eliminated by BLSTM. Therefore, we just need to set $k_1$ to a small number, which not only saves the computational load greatly but also is still effective in reducing the random noise as shown in Fig. \ref{fig:PCAvsMBN}b.
 
 Finally, $\delta$ controls how aggressively the nonlinearity of data is reduced. If the data is highly nonlinear, then we set $\delta$ to a large number which results in a very deep architecture that reduces the nonlinearity gradually layer-by-layer; otherwise, we set $\delta$ to a small number. As shown in Fig. \ref{fig:PCAvsMBN}a, the embedding vectors do not contain much nonlinearity. Therefore, we simply build MBN with a single hidden layer by setting $\delta=0$.

\section{SPEAKER SEPARATION EXPERIMENTS}

\subsection{Datasets}

\begin{figure}[t]
 \centering
 \begin{minipage}[t]{0.5\linewidth}
  \includegraphics[width=40mm]{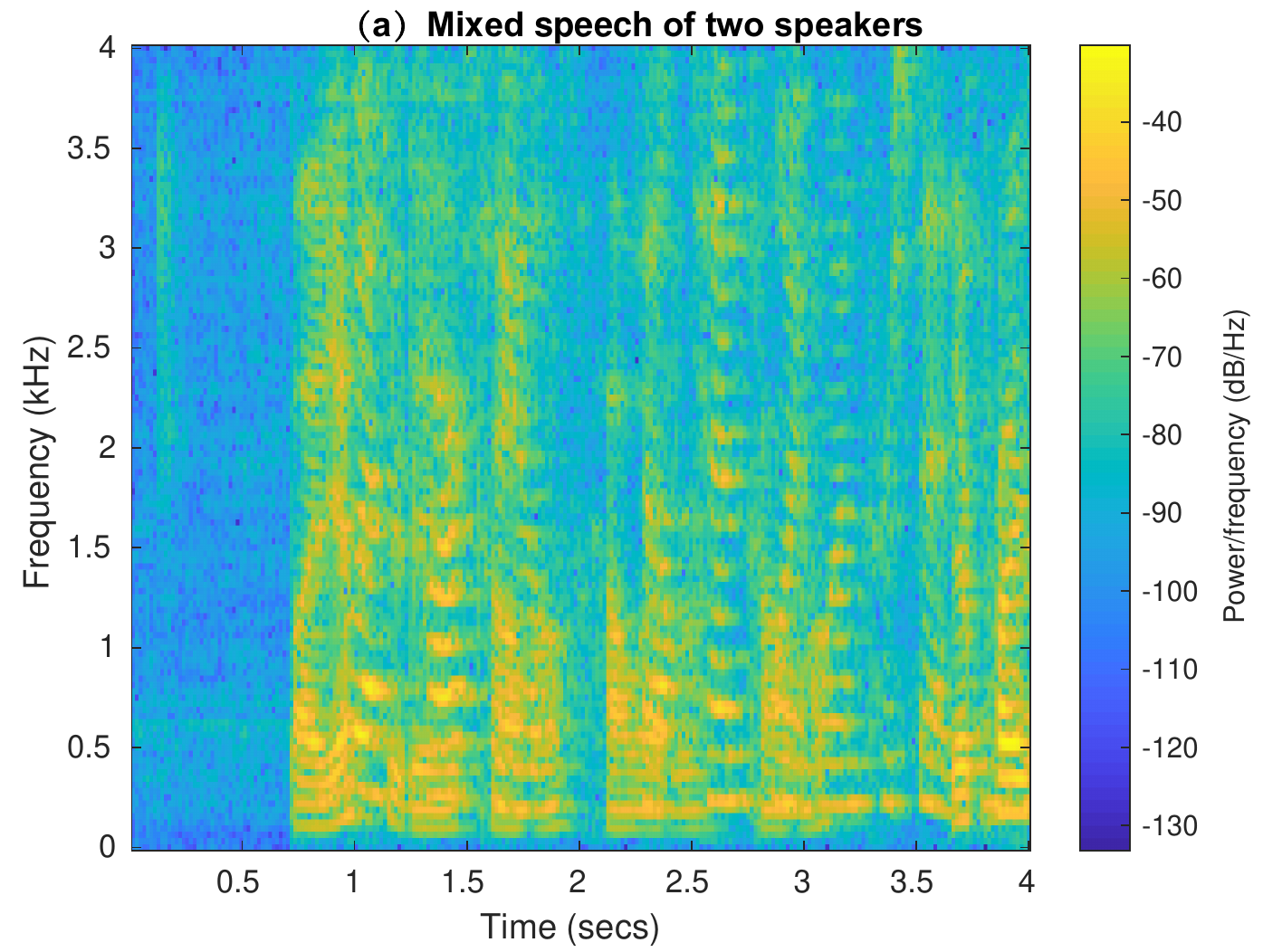}
  \end{minipage}%

  \centering
\begin{minipage}[t]{0.5\linewidth}
\includegraphics[width=40mm]{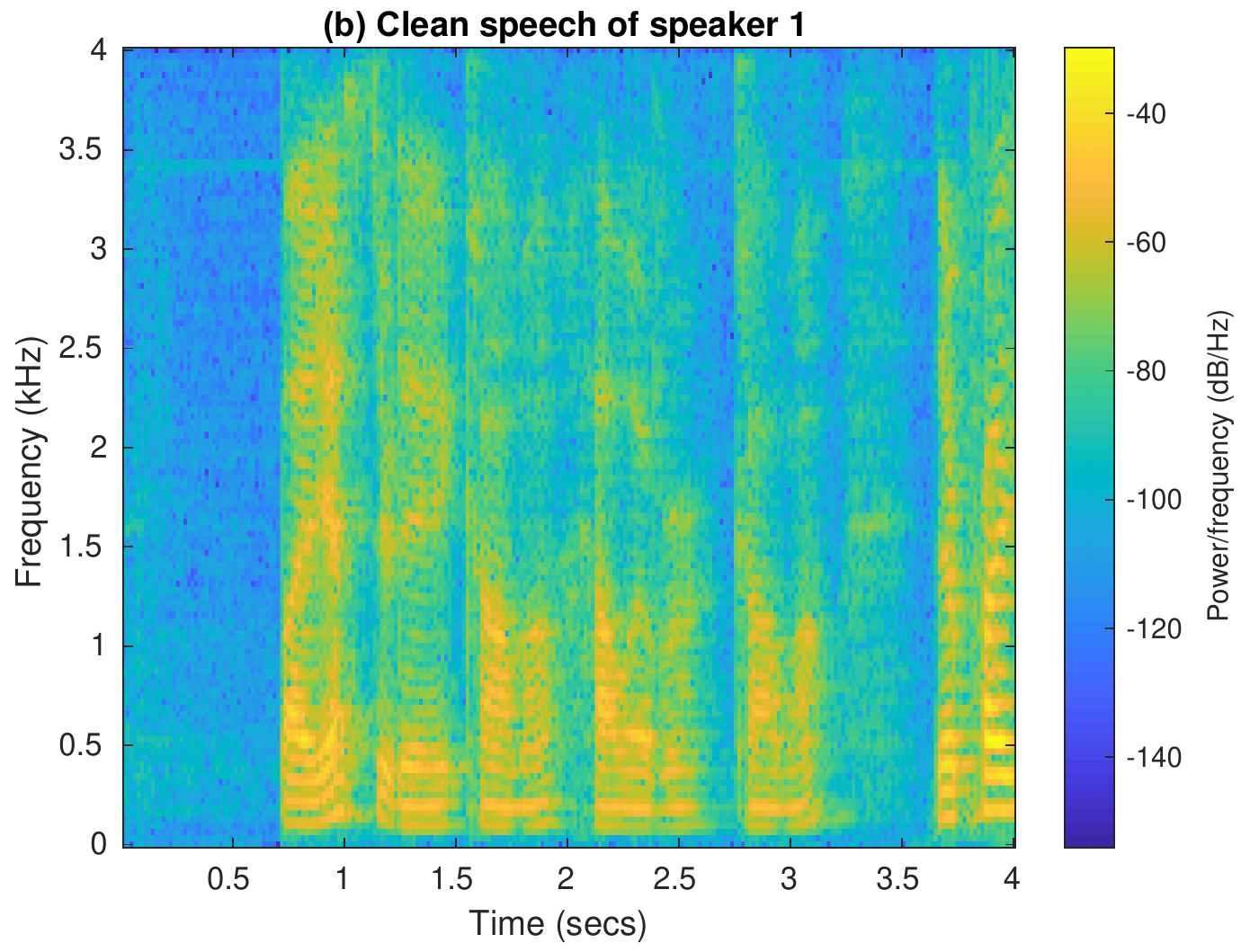}
  \label{fig:side:b}
\end{minipage}%
\begin{minipage}[t]{0.5\linewidth}
\centering
\includegraphics[width=40mm]{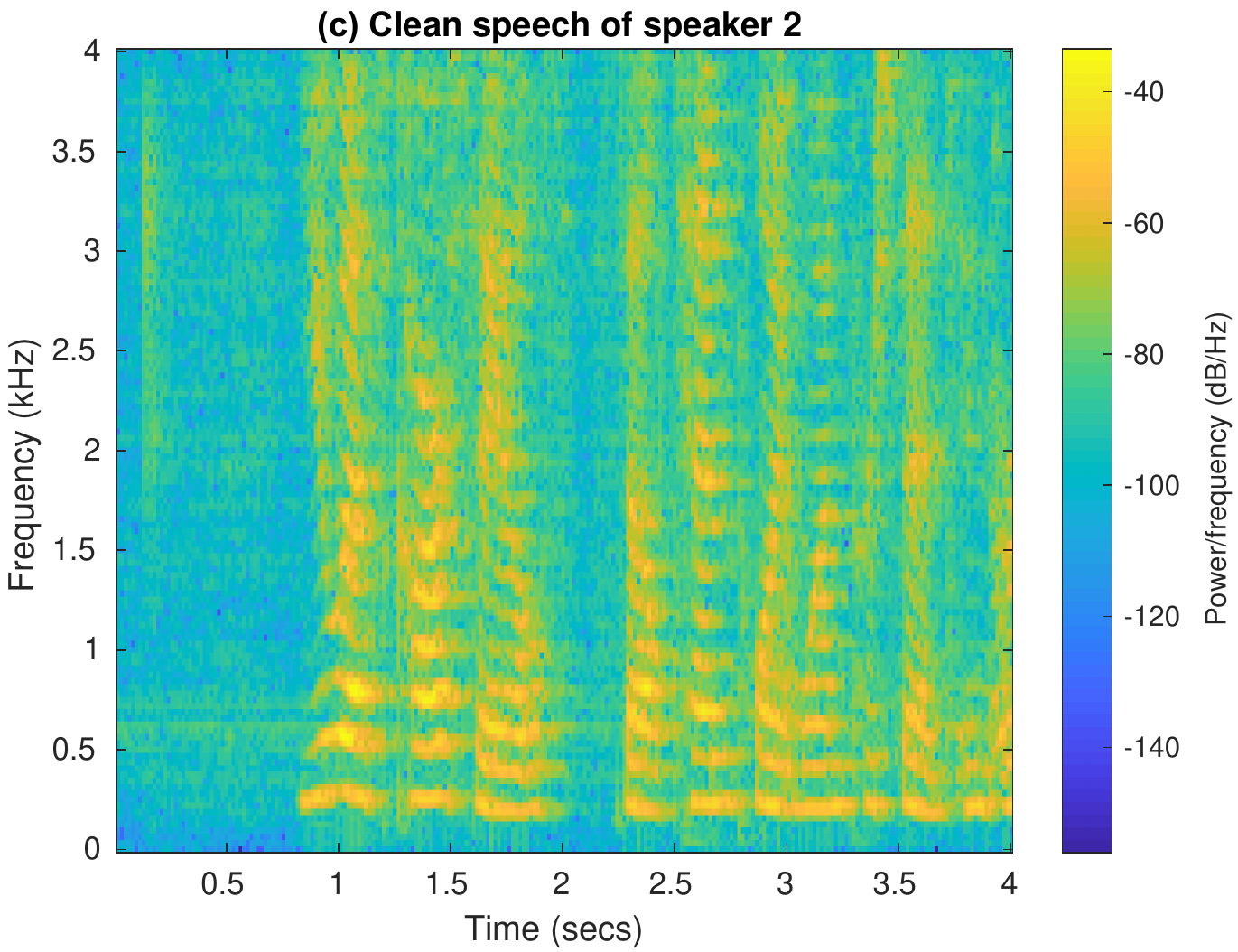}
  \label{fig:side:a}
 \end{minipage}
 
 \centering
  \begin{minipage}[t]{0.5\linewidth}
\includegraphics[width=40mm]{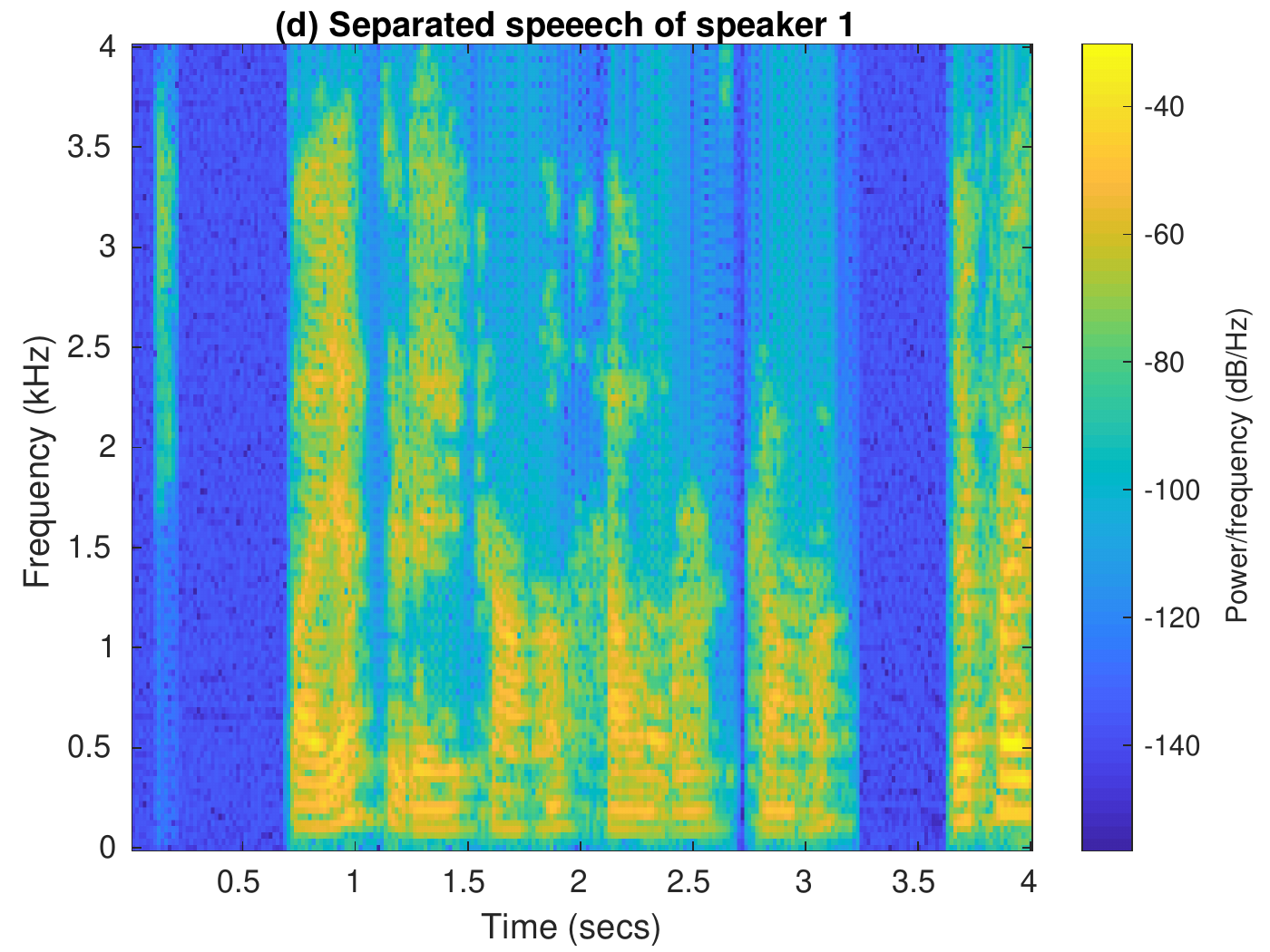}
  \label{fig:side:a2}
 \end{minipage}
 \begin{minipage}[t]{0.49\linewidth}
\includegraphics[width=40mm]{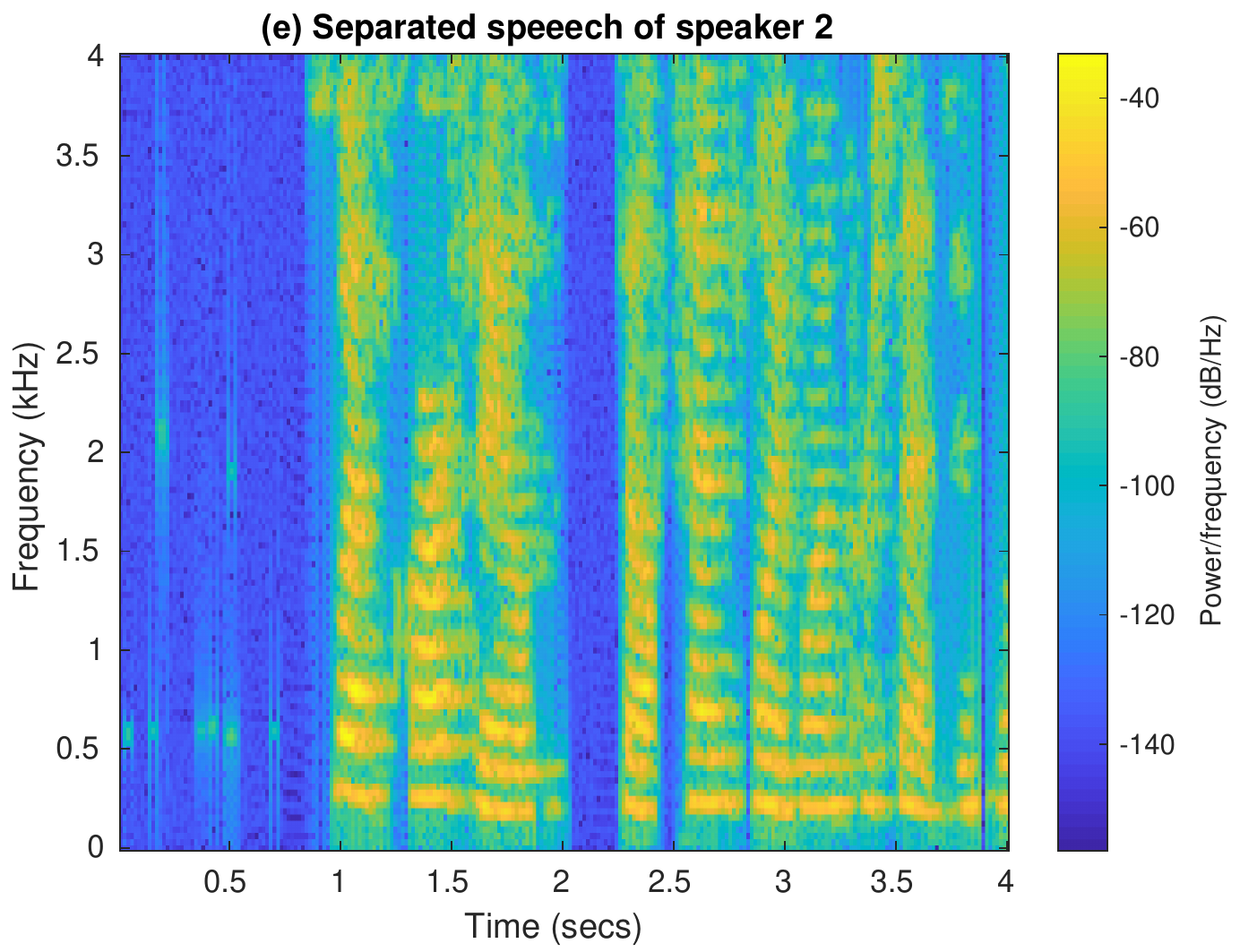}
  \label{fig:side:a3}
 \end{minipage}

  \caption{Logarithmic magnitude spectra of a mixed speech signal, its ground-truth components, and the estimated components produced by DPCL++ in the anechoic environment. (a) Mixed speech. (b) Clean speech of the first speaker. (c) Clean speech of the second speaker. (d) Estimated speech of the first speaker. (d) Estimated speech of the second speaker.}
  \label{fig:mix}
  \end{figure}

We used the WSJ0-2mix corpus as the speech source \cite{isik2016single,chen2017deep,luo2018speaker,Shi2019FurcaNeXt}, and resampled the speech data to 8 kHz. We focused on 2-speaker and 3-speaker speech separation problems. For the 2-speaker separation problem, we simulated both an anechoic environment and a reverberant environment for each mixture.
For each environment, we generated two datasets for the model training and test respectively. The training set contains 20000 mixtures. The test set contains 3000 mixtures. The two datasets are about 30 and 5 hours long respectively, which is enough to draw a reasonable experimental conclusion. To find the optimal hyperparameters, we further constructed a validation set containing 5000 mixtures.
For each mixture, we generated its anechoic recording by setting $T_{60}=0$, and its reverberant recording by selecting $T_{60}$ from a range of $[0.2, 0.6]$ second \cite{gannot2017consolidated}.
 Figures \ref{fig:mix}a to \ref{fig:mix}c show the log magnitude spectra of a mixture and its components in an anechoic environment. For the 3-speaker separation problem, we generated two test set of 3000 mixtures in the anechoic and reverberant environments respectively. We will evaluate the models trained for the 2-speaker separation problem on the 3-speaker test datasets directly.

 \begin{table*}[htbp]
\centering
   \caption{{Performance comparison between DPCL and DPCL++ in various experimental settings. The result of 1-channel DPCL was directly copied from \cite{hershey2016deep}.}}
   \label{tab:anechoic}
\begin{tabular}{lcccrrr}
\hline
\hline
Method&Number of speakers& Environment & Feature  &SDR (dB)&PESQ&STOI\\
\hline  
1-channel DPCL \cite{hershey2016deep}&2&anechoic &Log.mag&6.50&-&-\\
\hline
2-channel DPCL&2&anechoic&Log.mag+cosIPD&10.92&2.53&0.85\\
\textbf{2-channel DPCL++}&2&anechoic&Log.mag+cosIPD&\textbf{13.65}&\textbf{2.91}&\textbf{0.87}\\
\hline
2-channel DPCL&2&reverberant&Log.mag+cosIPD&8.61&2.28&0.73\\
\textbf{2-channel DPCL++}&2&reverberant&Log.mag+cosIPD&\textbf{10.70}&\textbf{2.51}&\textbf{0.75}\\
\hline
2-channel DPCL&3&reverberant&Log.mag+cosIPD&4.07&1.05&0.66\\
\textbf{2-channel DPCL++}&3&reverberant&Log.mag+cosIPD&\textbf{5.93}&\textbf{1.44}&\textbf{0.68}\\
\hline
\hline
\end{tabular}
\end{table*}
 
\subsection{Parameter Settings}

We set the frame length to 32 milliseconds and the frame shift to 8 milliseconds. We extracted a 129-dimensional Hamming window weighted STFT feature from each frame. We adopted a similar network structure of BLSTM with that in \cite{hershey2016deep}. Specifically, the BLSTM network consists of four hidden layers with 300 hidden units per layer. The network was optimized by stochastic gradient descent. The momentum was set to 0.9, and the learning rate was set to $10^{-5}$. To avoid falling into the local minima of BLSTM, we also added a Gaussian noise with a mean of 0 and a variance of 0.6 to the input. We evaluated the performance of the standard DPCL with the dimensions of the embedding vectors set to $D=\{10, 20, 40, 60\}$ respectively, and found that setting $D=40$ produced the best speech separation performance.
We set the hyperparameters of MBN as follows $V=400$, $a=0.9$, $k_1=20$, and $\delta=0$. We compared DPCL++ with DPCL \cite{hershey2016deep} given the same input acoustic features in multi-channel settings.

 \begin{table}[t]
\centering
   \caption{{ Effect of hyperparameter $\delta$ on performance in the anechoic environment.}}
   \label{tab:parameter}
\begin{tabular}{ccccc}
\hline
\hline
$\delta$&0.7& 0.5 & 0.3 & $\leq 0.1$\\
\hline  
SDR (dB) &8.81&10.86&12.57&13.65\\
\hline
\hline
\end{tabular}
\end{table}

The performance evaluation metrics include signal to distortion ratio (SDR) \cite{vincent2006performance}, perceptual evaluation of speech quality (PESQ) \cite{rix2001perceptual}, and short-time objective intelligibility (STOI) \cite{taal2011algorithm}. SDR is a metric similar to SNR for evaluating the quality of enhancement.  PESQ is a test methodology for automated assessment of the speech quality as experienced by a listener of a telephony system. STOI evaluates the objective speech intelligibility of time-domain signals. The higher the value of an evaluation metric is, the better the performance is.

\subsection{Results}

Figures \ref{fig:mix}d and \ref{fig:mix}e show the separation result of Fig. \ref{fig:mix}a. From the figure, we see that DPCL++ produces a good separation result close to its ground-truth. Table \ref{tab:anechoic} summarizes all comparison results. From the table, we see that DPCL++ achieves an SDR score of 2.73 dB higher than DPCL in the anechoic environment. It also achieves an SDR of 2.09 and 1.86 dB higher than DPCL in the 2-speaker and 3-speaker separation problems respectively in the reverberant environment. In addition, the PESQ score of DPCL++ is about 0.4 higher than that of DPCL, and the STOI score of DPCL++ is about 0.02 higher than DPCL on average. To summarize, DPCL++ outperforms DPCL in all experiments in terms of all three evaluation metrics.

\subsection{Effects of hyperparameters on performance}
 
 Due to the length limitation of this paper, we report the important effect of $\delta$ on performance in Table \ref{tab:parameter}. Because $k_1=20$, MBN builds a deep architecture when $\delta > 0.15$, and builds a shallow architecture when $\delta\leq 0.15$ according to \eqref{eq:k1} to \eqref{eq:kL}. From Table \ref{tab:parameter}, we see that building a shallow architecture achieves the best performance, while building a deep model degrades the performance on the contrary. Hence, the MBN with a single nonlinear layer not only helps improve the performance of DPCL++ but also saves a lot of computation load.

\section{CONCLUSIONs}

In this paper, we have proposed a multi-channel speaker-independent speech separation system, named DPCL++. It first extracts the cosIPD and log magnitude acoustic features as the input of BLSTM. Then, it produces an embedding vector for each T-F unit by BLSTM. Finally, it reduces the dimension of the embedding vectors by MBN, and uses the output of MBN for clustering. The core contribution of this paper is that DPCL++ introduces MBN into speech separation, which can be easily implemented. 
 We have compared DPCL++ with DPCL on the 2-speaker and 3-speaker speech separation problems in both the anechoic and reverberant environments.  
 Experimental results show that DPCL++ significantly outperforms DPCL in all test scenarios.


%
%
%


\small
\bibliographystyle{IEEEbib}
\bibliography{refs}

\end{document}